\documentclass[10pt, aps, pra, twocolumn, superscriptaddress, notitlepage]{revtex4-2}
\usepackage{graphicx}
\usepackage{CJK}
\usepackage{subfigure} 
\usepackage{float}
\usepackage{amsmath}
\usepackage{amssymb}
\usepackage{url}
\usepackage{natbib}

\usepackage[colorlinks,
			linkcolor=blue,
			anchorcolor=blue,
			citecolor=blue,
			urlcolor=blue]
{hyperref}    
\begin{document}
\title{Laser power stabilization using conservation law in acoustic optic modulator}
\author{Erwei Li}
\affiliation{National Time Service Center, Chinese Academy of Sciences, Xi'an, 710600 Shaanxi, China.}
\affiliation{University of Chinese Academy of Sciences, 100049 Beijing, China.}
\author{Qianjin\ Ma}
\affiliation{National Time Service Center, Chinese Academy of Sciences, Xi'an, 710600 Shaanxi, China.}
\author{Weiyu\ Wang}
\affiliation{National Time Service Center, Chinese Academy of Sciences, Xi'an, 710600 Shaanxi, China.}
\affiliation{University of Chinese Academy of Sciences, 100049 Beijing, China.}
\author{Bobo\ Du}
\email{bobo.du@xjtu.edu.cn}
\affiliation{School of Electronic Science and Engineering, Xi'an Jiaotong University, Xi'an, 710049 Shaanxi, China.}
\author{\ Guobin\ Liu}
\email{liuguobin@ntsc.ac.cn}
\affiliation{National Time Service Center, Chinese Academy of Sciences, Xi'an, 710600 Shaanxi, China.}
\affiliation{University of Chinese Academy of Sciences, 100049 Beijing, China.}
\affiliation{Key Laboratory of Time Reference and Applications, Chinese Academy of Sciences, Xi'an, 710600 Shaanxi, China.}
\date{\today}

\begin{abstract}
Laser power stabilization plays an important role in modern precision instruments based on atom-laser interactions. Here we demonstrate an alternative active control method of laser power utilizing the conservation law in an acoustic optic modulator (AOM). By adjusting the 1st order beam power to dynamically follow the fluctuation of the total power of all diffraction beams, the 0th order application beam as the difference term, is stabilized. Experimental result demonstrates that the relative power noise of the controlled application beam is reduced by a factor of 200, reaching $4 \times 10^{-6} $ Hz$^{-1/2}$ at 10$^{-4}$ Hz compared with the uncontrolled total power. Allan deviation shows that the application beam reaches a relative power instability of 3.28$\times 10^{-6}$ at 500 s averaging time. In addition, the method allows a high availability of total power source. The method opens a new way of laser power stabilization and shall be very useful in applications such as atomic clocks, laser interferometers and gyroscopes.
\end{abstract}
\pacs{}
\maketitle

Stable laser power sources play significant role in the long-term performance of precision measurement physical instruments such as atomic clocks, interferometers and gyroscopes. As well known, AC Stark shift, the so-called light shift contributes as one major source of systematic errors in various kinds of laser pumped atomic clocks from microwave to optical frequency range, and makes the clock's frequency stability depending significantly on the laser power stability\cite{Vanier2007,Ludlow2015,Katori2015,Kitching2020,Mileti2020}. Lots of efforts have been made to reduce the laser power noise in the famous Laser Interferometer Gravitational-Wave Observatory (LIGO) experiments\cite{Seifert2006,Kwee2008,Kwee2009,Kwee2012,Junker2017,Vahlbruch2018} and Laser Interferometer Space Antenna (LISA) projects\cite{LISA2022,LISA2023}. In fiber optics sensors such as the fiber-optic gyroscopes, laser power fluctuations can decisively affect measurement accuracy\cite{FOG1994,FOS2023}.

Laser power stabilization is usually realized through an active control loop including a power regulator such as electro-optic or acoustic-optic modulators\cite{Robertson1986,Tran1993,Seifert2006,Kwee2008,Kwee2009,Kwee2012,Junker2017,Vahlbruch2018}. There are also methods using alternative optical elements as the power regulators in the control loop, such as optical AC coupling cavity \cite{Kaufer2019}, micro-vibration mirror \cite{TradNery2021}, photosensitive lens \cite{Guan2021} and non-polarized beam splitter (NPBS) \cite{Wang2021}. Until now, the state of the art record for laser power stabilization was realized by the Max-Planck-Institution's laser power stabilization apparatus for LIGO, which achieved a relative power noise (RPN) of $1.8\times10^{-9}$ Hz$^{-1/2}$ in the frequency range from 100 Hz to 1 kHz\cite{Junker2017}. As for the long-term laser power stabilization, researchers from Paris Observatory achieved an Allan deviation $2\times10^{-6}$ for an averaging time up to $10^{4}$ seconds\cite{Tricot2018}. However, modern precision measurement instruments using atom-laser interactions are typically built to be complex and large table-top systems, it is therefore difficult to realize high accuracy temperatures control as was done within a relatively small size setup \cite{Tricot2018}. Additionally, in above laser power control loops, the long-term stability of the application beam is usually inferior to that of the control beam due to the deviation of the beam splitter's splitting ratio from optimized value \cite{Tricot2018,Wang2021}. Optimized splitting ratio of the beam splitter also limits the availability of application beam power. From the perspective of control methodology, the long-term stability becomes deteriorated because the application beam was not deeply engaged in the control loop.
\begin{figure*}
	\centering
	\includegraphics[width=0.85\textwidth]{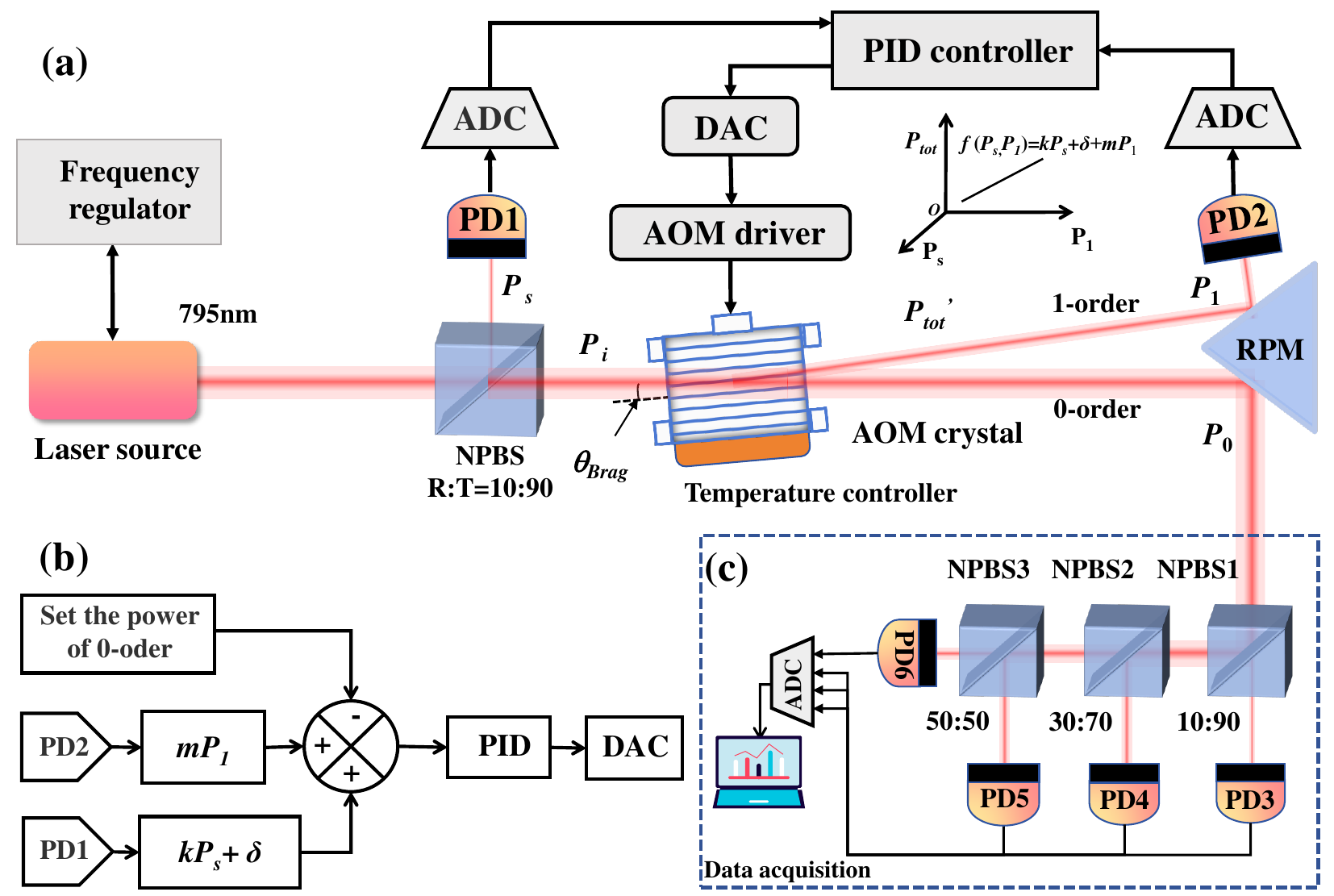}
	\caption{Schematic diagram of the laser power stabilization method. (a) Experimental setup. The laser source beam is split by non-polarized beam splitter (NPBS). The reflection beam power $P_s$ is collected by photodetector PD1. The transmission beam hits on the acoustic optic modulator (AOM) input as its incidence beam with power $P_i$. The AOM is temperature-controlled. The 0th and 1st order diffraction beams are reflected by a right-angle prism mirror (RPM), then their beam power $P_1$ and $P_0$ is collected by photodetectors PD2 and PD3-PD6 in (c) respectively. The higher-order diffraction beams are blocked. The conservation law can be represented by the function in the Cartesian coordinates, $P_{tot}=k P_s+\delta+m P_1$, saying the total power proportional to the sampling beam power and the 1st order beam power, plus a constant attenuation term. The PID controller takes the $P_s$ and $P_1$ as inputs and outputs a calibrated voltage to adjust the AOM diffraction efficiency via the AOM driver. The calibration voltage is generated by the (b) control algorithms designed according to the conservation law.}
	\label{fig1}
\end{figure*} 

In this work, we realized a long-term laser power stabilization using the conservation law in an AOM. The conservation law can be briefly described as: the total output power of AOM equals to the sum power of the 0th, the 1st order diffraction beams and a temperature-sensitive term. The powers of the 0th and 1st order diffraction beams can be characterized by that of a sampling beam and the temperature-sensitive term is found to be proportional to the power of the 1st order diffraction beam. Utilizing the conservation law, we design an algorithm with a digital proportional-integral-differential (PID) controller. The control method enables dynamical adjustment of the 1st order beam power according to the total power fluctuations in real time. By maintaining a constant difference between the power of the total diffraction beams and the 1st order diffraction beam of the AOM, the power of the 0th order diffraction beam as the application beam is stabilized significantly. A relative power noise of $4\times10^{-6}$ Hz$^{-1/2}$ at $10^{-4}$ Hz and an Allan deviation of 3.28$\times 10^{-6}$ at 500 s averaging time are achieved. This work opens a totally new way of laser power stabilization and shall be very easily adapted to various applications where long term stability of laser power is highly demanded.

Our experimental setup is depicted in Fig. \ref{fig1}. The photoelectric control system is shown in Fig. \ref{fig1}(a). A semiconductor laser (EYP-DFB-0795-00080-1500-BFW01-0005, TOPTICA) is used as the source. About 0.5 mW laser power is feed into the frequency regulator, which is used for locking the laser frequency to specific atomic transition line. The major beam power incidents into the NPBS with reflection to transmission ratio R:T=10:90. The transmitted beam goes into the AOM (model 3080-125, Gooch-Housego) at the Bragg angle, while the reflected beam is sampled by photodetector PD1 (PDA36A2, Thorlabs). The diffracted beams contains the 0th, 1st and higher-order beams. The 0th and 1st order beams exhibit insufficient spacing for direct measurement. To resolve this constraint, the optical path is extended. Subsequently, a right-angle prism mirror (RPM) is implemented to redirect the beams prior to measurement, while the higher-order beams are blocked. The 1st order beam is detected by photodetector PD2. As the high power of the 0th order beam can saturate a single photodetector, three NPBS and four photodetectors are used to split the beam and collect the beam powers accordingly, as shown in Fig. \ref{fig1}(c). The detected voltages are digitized by an analog-to-digital converter (ADC) and subsequently processed through a data acquisition (DAQ) to estimate the results of power control. Since the diffraction efficiency of the AOM is sensitive to temperature fluctuations \cite{Kobayashi2006,Zhang2019,Deng2023}, a TEC cooling plate and an NTC thermistor are used to control the temperature of the AOM housing to 23.02±0.006 $^{\rm o}$C. Photodetector PD1 measures the sampling beam power and characterizes the AOM incidence power. Photodetector PD2 measures the 1st order diffraction beam power and characterizes the temperature-sensitive term. In Fig. \ref{fig1}(b), according to the set value of the 0th order beam power, a PID control algorithm is utilized to make the AOM driver adjusting the 1st order diffraction beam of the AOM. 

\begin{figure*}
	\centering
	\includegraphics[width=0.75\textwidth]{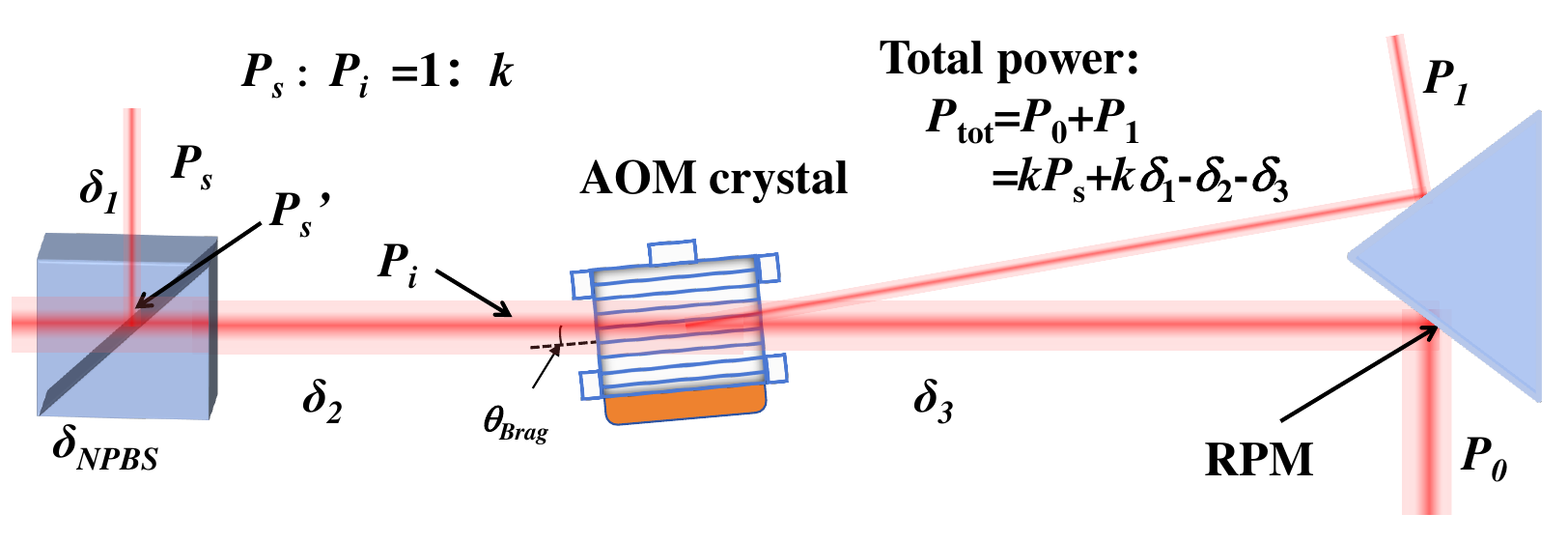}
	\caption{Schematic diagram showing the conservation law of the optical power distribution in the main optical path.}
	\label{fig2}
\end{figure*}

The control algorithm are designed based on the conservation law, which is revealed by a thorough analysis on the power distribution in optical path before and after the AOM, as shown in Fig. \ref{fig2}. First, the sampling beam power $P_{s}$ and the power $P_s^{'}$ at the splitting interface of NPBS is given by
\begin{equation}
P_s = P_s^{'} - \delta_1,
\label{eq1}
\end{equation}
where $\delta_1$ is the attenuation caused by the NPBS and the air from the reflected point to PD1. The AOM incidence power $P_{i}$ is given by
\begin{equation}
	P_i = kP_s' - \delta_2,
	\label{eq2}
\end{equation}
where $k$ is the transmission to reflection ratio of NPBS, and $\delta_2$ is the attenuation in between the splitting interface of NPBS and the incidence point of AOM. The total output power $P_{tot}$ of the AOM can be given by
\begin{equation}
	P_{tot} = P_i - \delta_3,
	\label{eq3}
\end{equation}
where $\delta_3$ is the attenuation of incidence beam in the acousto-optic crystal and in the air. According to Eq. \ref{eq1}-\ref{eq3}, the total power at the measured position can be obtained
\begin{equation}
	\begin{split}
	P_{tot} &= P_0 + P_1 \\&= kP_s + k\delta_1 - \delta_2 - \delta_3 \\&= kP_s + \delta.
	\end{split}
	\label{eq4}
\end{equation}
After establishing stable optical paths and components, the total attenuation will stay constant. We therefore define the composite attenuation constant as: $\delta = k\delta_1-\delta_2-\delta_3$. Since the diffraction efficiency in this experiment remains extremely low ($< 2\%$), the power of higher order beams can be safely neglected in this case.

\begin{figure}
	\centering
	\includegraphics[width=0.48\textwidth]{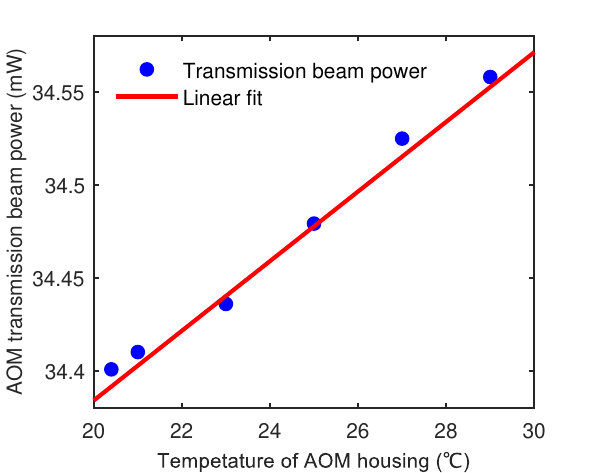}
	\caption{The transmission beam power measured as a function of the AOM housing temperature.}
	\label{fig3}
\end{figure}

During experiments, we found the transmission of the AOM crystal increases with the temperature. To test the AOM's temperature dependence, we turned off the frequency regulator and maintain a fixed driving current of laser source to ensure a constant beam power incident on the AOM. By turning off the AOM driver, we set the diffraction efficiency to zero and then adjust the housing temperature of the AOM, the transmission beam power versus the temperature is plotted in Fig. \ref{fig3}. It can be seen that the change of the transmission beam power with temperature is approximately linear, with a slope rate of about 0.018 mW/$^{\circ}$C.

\begin{figure*}
\centering
\includegraphics[width=1.0\textwidth]{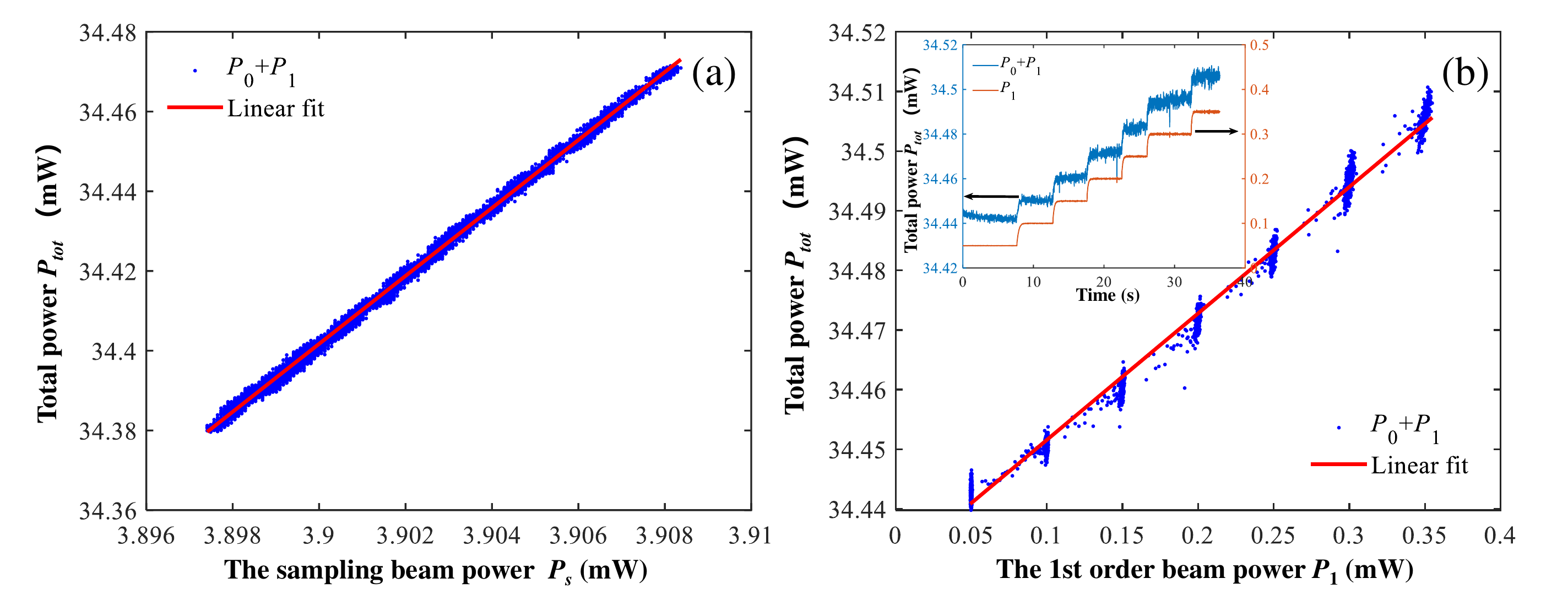}
\label{fig4}
\caption{(a) The relationship between the total diffraction beam power $(P_0+P_1)$ and the sampling beam power $P_s$. (b) The relationship between the total diffraction beam power $(P_0+P_1)$ and the 1st order beam power $P_1$. }
\label{fig4}
\end{figure*}

On the other hand, we noticed that the temperature of AOM crystal increases with the AOM driver output voltage \cite{Zhang2019,Kobayashi2006}. Once the AOM is turned on, there should be an additional temperature-sensitive power fluctuation due to the heating effect of the nonzero AOM driver output. To address this issue quantitatively, we introduce a temperature-sensitive power term $P_{T}$ into Eq. \ref{eq4}, which is now modified to be
\begin{equation}
P_{tot} = P_0 + P_1 = kP_s + \delta + P_T.
\label{eq5}
\end{equation}
When the diffraction efficiency is at a low level ($< 3\%$), the temperature-sensitive power term is proportional to the AOM driver output voltage and thus the 1st order diffraction beam power, therefore Eq. \ref{eq5} can be rewritten as
\begin{equation}
	P_{tot} = P_0 +P_1 = kP_s + \delta + mP_1,
	\label{eq6}
\end{equation}
where $m$ is a unit-less coefficient of the temperature-sensitive power over the 1st order diffraction beam power $P_1$. 

To this end, we have derived the relationship between the 0th order diffraction beam power, the 1st order diffraction beam power and the sampling beam power from a theoretical perspective according to the power distribution analysis in Fig. \ref{fig2}. Furthermore, we have to prove this so-called conservation law experimentally. By tuning the driving current of the laser source, we measured the powers of the sampling beam, the 1st order diffraction beam and the 0th order diffraction beam with the photodetectors PD1, PD2 and PD3-PD6, respectively. The total power $P_{tot}$ changes with the sampling beam power $P_s$ and the 1st order diffraction beam power $P_1$ approximately linearly, as shown in Fig. \ref{fig4}(a) and \ref{fig4}(b), respectively. This proves the relationship in Eq. \ref{eq6} is correct and gives us the values of coefficients $k=8.5301$, $m=0.208$ and the value of constant attenuation $\delta=1.1345$ mW through linear fittings.

Since all the power terms fluctuate with time, we can rewrite Eq. \ref{eq6} as
\begin{equation}
	P_0(t) = kP_s(t) + (m-1) P_1(t) +\delta.  
	\label{eq7}
\end{equation}
As $m$ is smaller than one, the second term in Eq. \ref{eq7} is actually negative. In order to keep the application beam power $P_0(t)$ stable over time, we need to keep the difference between the total power $P_{tot}$ given by Eq. \ref{eq6} and the 1st order diffraction beam power $P_1$ dynamically constant. This is done by the PID controller. It takes the power of the sampling beam and the 1st order beam (in Fig. \ref{fig1}(a)) as inputs, and outputs a calibrated AOM driver control voltage to adjust the diffraction beam power. The control voltage of AOM driver is calibrated using an algorithm depicted in Fig. \ref{fig1}(b). The basic idea is to adjust the power of the 1st order beam in real time to follow the power of the sampling beam dynamically, as described by Eq. \ref{eq7}.

\begin{figure}
	\centering
	\includegraphics[width=0.48\textwidth]{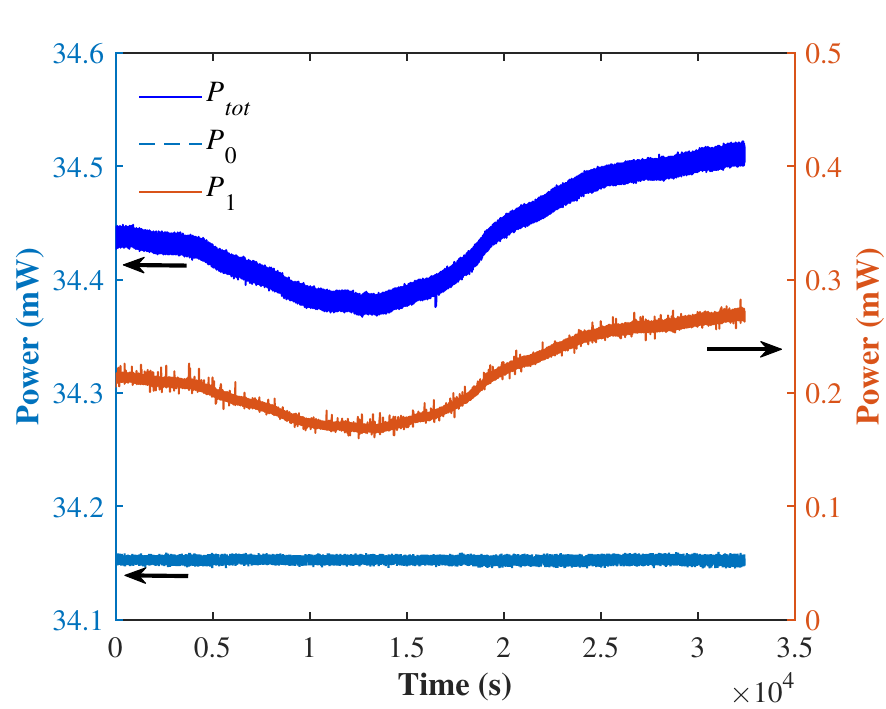}
	\caption{Temporal fluctuations of the total power $P_{tot}$, the 1st order beam power $P_1$ and the 0th order beam power $P_0$ during the period of 9 hours.}
	\label{fig5}
\end{figure}

During about 9 hours continuous operation, the total power, the power of the 1st and 0th order beams are recorded and plotted in Fig. \ref{fig5}. It is clear that the 1st order beam power follows the fluctuation of the total power in real time and the power of the application (the 0th order) beam stays stable at a level of $\sim$0.01 mW at a baseline of 34.15 mW. Surprisingly, the control is successful while the control beam (the 1st order beam) takes only 1$\%$ and the application beam (the 0th order beam) take 99$\%$ of the total power.

\begin{figure}
	\centering
	\includegraphics[width=0.48\textwidth]{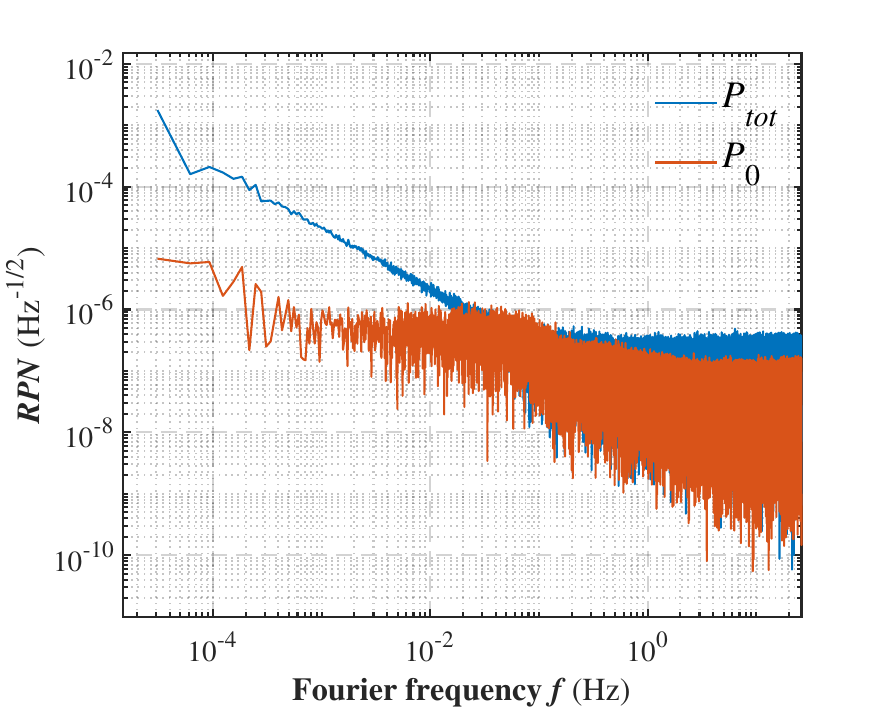}
	\caption{Relative power noise of the uncontrolled total power $P_{tot}$ and the controlled application beam $P_0$  in the upper trace and lower trace, respectively.} 
	\label{fig6}
\end{figure}

\begin{figure}
	\centering
	\includegraphics[width=0.48\textwidth]{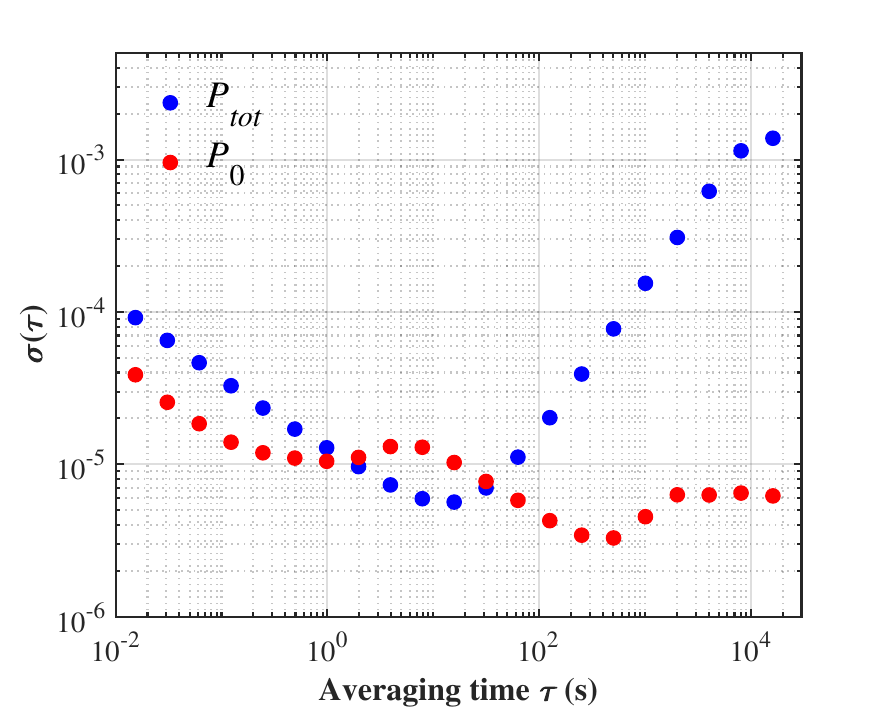}
	\caption{Allan deviation of the uncontrolled total power $P_{tot}$ and controlled application beam power $P_0$ in the upper trace and lower trace, respectively.}
	\label{fig7}
\end{figure}

To test the control method's performance, the relative power noise of the total power $P_{tot}$ and the application beam power $P_0$ recorded in Fig. \ref{fig5} is shown as the upper and lower lines respectively in Fig. \ref{fig6}. Compared to the uncontrolled total power, there is a notably reduction of RPN for the controlled application beam at the frequency range from $6\times10^{-4}$ Hz up to 0.02 Hz. The RPN is reduced by a factor of 200, 20 and 5, reaching $4 \times 10^{-6} $ Hz$^{-1/2}$ at 0.0001 Hz, $1 \times 10^{-6} $ Hz$^{-1/2}$ at 0.001 Hz and $1 \times 10^{-6} $ Hz$^{-1/2}$ at 0.01 Hz, respectively. There is also some RPN reduction even at high frequency above 1 Hz. Moreover, the Allan deviation of the data is shown in Fig. \ref{fig7}. The long term stability of the controlled application beam power is better than that of the uncontrolled total power at averaging time ranging from 0.1 to 1 second and from 30 seconds to 4.48 hours.

Within the averaging time 1 to 30 s (with respect to a bit RPN increasing at the 0.03-0.1 Hz frequency range), there is some notable deterioration of the Allan deviation. It indicates uncontrolled residual noises. We suspect three possible noise sources: 1) vibration of the optical parts such as the AOM, NPBS and RPM. 2) thermal gradient of the AOM crystal. The ultrasonic wave can introduce temperature gradients across the diffraction crystal, causing the directional drift of the 1st order beam \cite{Zhang2019,Kobayashi2006} and leading to beam jitter incident on the photodetector. 3) uncontrolled attenuation. We have assumed a constant attenuation $\delta$ throughout the work. However, there are probably some variations of these attenuation due to the performance fluctuations of the crystals and detectors. Our method is based on a thorough analysis of beam power distribution within the AOM control loop, assisted with the developed digital PID algorithm. In this case, further detailed analysis on the properties of aforementioned noise sources assisted with more intelligent modern control algorithms (for example, neural networks) should reduce these noise terms. One point to note is that, our relative power noise analysis can be done only at frequency up to about 30 Hz for now. This is due to the limited execution rate of the PID controller and it will be resolved by an update in the future. 

In conclusion, a laser power stabilization method is realized utilizing the conservation law of AOM regulator. The conservation law is derived theoretically and proved effective to the first order by experiment. The method uses only a little portion of power (1$\%$) for controlling and leaves most (99$\%$) of the total power for the application beam. Despite such little power used by the control beam, the method achieves significant improvement on the long term stability of the controlled application beam power over the uncontrolled total power. Experimental results show that the relative power noise of the controlled application beam is reduced by a factor of 200, reaching $4 \times 10^{-6} $ Hz$^{-1/2}$ at 0.0001 Hz compared with the uncontrolled total power. Allan deviation shows that the application beam power reaches a relative instability of 3.28$\times 10^{-6}$ at 500 s averaging time and increase to 6.19$\times 10^{-6}$ at 4.48 hour averaging time. This method also opens a totally new way for laser power stabilization and further improvement is possible in the future. The method is potentially useful in areas such as atomic clocks, laser interferometers and gyroscopes, which require the laser power ultrastable in the long-term operations.

The author Guobin Liu appreciate the financial support of Chinese Academy of Sciences under Grant No. E209YC1101.


\begin{thebibliography}{25}%
\makeatletter
\providecommand \@ifxundefined [1]{%
 \@ifx{#1\undefined}
}%
\providecommand \@ifnum [1]{%
 \ifnum #1\expandafter \@firstoftwo
 \else \expandafter \@secondoftwo
 \fi
}%
\providecommand \@ifx [1]{%
 \ifx #1\expandafter \@firstoftwo
 \else \expandafter \@secondoftwo
 \fi
}%
\providecommand \natexlab [1]{#1}%
\providecommand \enquote  [1]{``#1''}%
\providecommand \bibnamefont  [1]{#1}%
\providecommand \bibfnamefont [1]{#1}%
\providecommand \citenamefont [1]{#1}%
\providecommand \href@noop [0]{\@secondoftwo}%
\providecommand \href [0]{\begingroup \@sanitize@url \@href}%
\providecommand \@href[1]{\@@startlink{#1}\@@href}%
\providecommand \@@href[1]{\endgroup#1\@@endlink}%
\providecommand \@sanitize@url [0]{\catcode `\\12\catcode `\$12\catcode
  `\&12\catcode `\#12\catcode `\^12\catcode `\_12\catcode `\%12\relax}%
\providecommand \@@startlink[1]{}%
\providecommand \@@endlink[0]{}%
\providecommand \url  [0]{\begingroup\@sanitize@url \@url }%
\providecommand \@url [1]{\endgroup\@href {#1}{\urlprefix }}%
\providecommand \urlprefix  [0]{URL }%
\providecommand \Eprint [0]{\href }%
\providecommand \doibase [0]{https://doi.org/}%
\providecommand \selectlanguage [0]{\@gobble}%
\providecommand \bibinfo  [0]{\@secondoftwo}%
\providecommand \bibfield  [0]{\@secondoftwo}%
\providecommand \translation [1]{[#1]}%
\providecommand \BibitemOpen [0]{}%
\providecommand \bibitemStop [0]{}%
\providecommand \bibitemNoStop [0]{.\EOS\space}%
\providecommand \EOS [0]{\spacefactor3000\relax}%
\providecommand \BibitemShut  [1]{\csname bibitem#1\endcsname}%
\let\auto@bib@innerbib\@empty
\bibitem [{\citenamefont {Vanier}\ and\ \citenamefont
  {Mandache}(2007)}]{Vanier2007}%
  \BibitemOpen
  \bibfield  {author} {\bibinfo {author} {\bibfnamefont {J.}~\bibnamefont
  {Vanier}}\ and\ \bibinfo {author} {\bibfnamefont {C.}~\bibnamefont
  {Mandache}},\ }\bibfield  {title} {\bibinfo {title} {The passive optically
  pumped rb frequency standard: the laser approach},\ }\href@noop {} {\bibfield
   {journal} {\bibinfo  {journal} {Applied Physics B}\ }\textbf {\bibinfo
  {volume} {87}},\ \bibinfo {pages} {565} (\bibinfo {year} {2007})}\BibitemShut
  {NoStop}%
\bibitem [{\citenamefont {Ludlow}\ \emph {et~al.}(2015)\citenamefont {Ludlow},
  \citenamefont {Boyd}, \citenamefont {Ye}, \citenamefont {Peik},\ and\
  \citenamefont {Schmidt}}]{Ludlow2015}%
  \BibitemOpen
  \bibfield  {author} {\bibinfo {author} {\bibfnamefont {A.~D.}\ \bibnamefont
  {Ludlow}}, \bibinfo {author} {\bibfnamefont {M.~M.}\ \bibnamefont {Boyd}},
  \bibinfo {author} {\bibfnamefont {J.}~\bibnamefont {Ye}}, \bibinfo {author}
  {\bibfnamefont {E.}~\bibnamefont {Peik}},\ and\ \bibinfo {author}
  {\bibfnamefont {P.~O.}\ \bibnamefont {Schmidt}},\ }\bibfield  {title}
  {\bibinfo {title} {Optical atomic clocks},\ }\href
  {https://doi.org/10.1103/RevModPhys.87.637} {\bibfield  {journal} {\bibinfo
  {journal} {Rev. Mod. Phys.}\ }\textbf {\bibinfo {volume} {87}},\ \bibinfo
  {pages} {637} (\bibinfo {year} {2015})}\BibitemShut {NoStop}%
\bibitem [{\citenamefont {Katori}\ \emph {et~al.}(2015)\citenamefont {Katori},
  \citenamefont {Ovsiannikov}, \citenamefont {Marmo},\ and\ \citenamefont
  {Palchikov}}]{Katori2015}%
  \BibitemOpen
  \bibfield  {author} {\bibinfo {author} {\bibfnamefont {H.}~\bibnamefont
  {Katori}}, \bibinfo {author} {\bibfnamefont {V.~D.}\ \bibnamefont
  {Ovsiannikov}}, \bibinfo {author} {\bibfnamefont {S.~I.}\ \bibnamefont
  {Marmo}},\ and\ \bibinfo {author} {\bibfnamefont {V.~G.}\ \bibnamefont
  {Palchikov}},\ }\bibfield  {title} {\bibinfo {title} {Strategies for reducing
  the light shift in atomic clocks},\ }\href
  {https://doi.org/10.1103/PhysRevA.91.052503} {\bibfield  {journal} {\bibinfo
  {journal} {Phys. Rev. A}\ }\textbf {\bibinfo {volume} {91}},\ \bibinfo
  {pages} {052503} (\bibinfo {year} {2015})}\BibitemShut {NoStop}%
\bibitem [{\citenamefont {Abdel~Hafiz}\ \emph {et~al.}(2020)\citenamefont
  {Abdel~Hafiz}, \citenamefont {Vicarini}, \citenamefont {Passilly},
  \citenamefont {Calosso}, \citenamefont {Maurice}, \citenamefont {Pollock},
  \citenamefont {Taichenachev}, \citenamefont {Yudin}, \citenamefont
  {Kitching},\ and\ \citenamefont {Boudot}}]{Kitching2020}%
  \BibitemOpen
  \bibfield  {author} {\bibinfo {author} {\bibfnamefont {M.}~\bibnamefont
  {Abdel~Hafiz}}, \bibinfo {author} {\bibfnamefont {R.}~\bibnamefont
  {Vicarini}}, \bibinfo {author} {\bibfnamefont {N.}~\bibnamefont {Passilly}},
  \bibinfo {author} {\bibfnamefont {C.}~\bibnamefont {Calosso}}, \bibinfo
  {author} {\bibfnamefont {V.}~\bibnamefont {Maurice}}, \bibinfo {author}
  {\bibfnamefont {J.}~\bibnamefont {Pollock}}, \bibinfo {author} {\bibfnamefont
  {A.}~\bibnamefont {Taichenachev}}, \bibinfo {author} {\bibfnamefont
  {V.}~\bibnamefont {Yudin}}, \bibinfo {author} {\bibfnamefont
  {J.}~\bibnamefont {Kitching}},\ and\ \bibinfo {author} {\bibfnamefont
  {R.}~\bibnamefont {Boudot}},\ }\bibfield  {title} {\bibinfo {title} {Protocol
  for light-shift compensation in a continuous-wave microcell atomic clock},\
  }\href {https://doi.org/10.1103/PhysRevApplied.14.034015} {\bibfield
  {journal} {\bibinfo  {journal} {Phys. Rev. Appl.}\ }\textbf {\bibinfo
  {volume} {14}},\ \bibinfo {pages} {034015} (\bibinfo {year}
  {2020})}\BibitemShut {NoStop}%
\bibitem [{\citenamefont {Almat}\ \emph {et~al.}(2020)\citenamefont {Almat},
  \citenamefont {Gharavipour}, \citenamefont {Moreno}, \citenamefont {Gruet},
  \citenamefont {Affolderbach},\ and\ \citenamefont {Mileti}}]{Mileti2020}%
  \BibitemOpen
  \bibfield  {author} {\bibinfo {author} {\bibfnamefont {N.}~\bibnamefont
  {Almat}}, \bibinfo {author} {\bibfnamefont {M.}~\bibnamefont {Gharavipour}},
  \bibinfo {author} {\bibfnamefont {W.}~\bibnamefont {Moreno}}, \bibinfo
  {author} {\bibfnamefont {F.}~\bibnamefont {Gruet}}, \bibinfo {author}
  {\bibfnamefont {C.}~\bibnamefont {Affolderbach}},\ and\ \bibinfo {author}
  {\bibfnamefont {G.}~\bibnamefont {Mileti}},\ }\bibfield  {title} {\bibinfo
  {title} {Long-term stability analysis toward 10-14 level for a highly compact
  pop rb cell atomic clock},\ }\href
  {https://doi.org/10.1109/TUFFC.2019.2940903} {\bibfield  {journal} {\bibinfo
  {journal} {IEEE Transactions on Ultrasonics Ferroelectrics and Frequency
  Control}\ }\textbf {\bibinfo {volume} {67}},\ \bibinfo {pages} {207}
  (\bibinfo {year} {2020})}\BibitemShut {NoStop}%
\bibitem [{\citenamefont {Seifert}\ \emph {et~al.}(2006)\citenamefont
  {Seifert}, \citenamefont {Kwee}, \citenamefont {Heurs}, \citenamefont
  {Willke},\ and\ \citenamefont {Danzmann}}]{Seifert2006}%
  \BibitemOpen
  \bibfield  {author} {\bibinfo {author} {\bibfnamefont {F.}~\bibnamefont
  {Seifert}}, \bibinfo {author} {\bibfnamefont {P.}~\bibnamefont {Kwee}},
  \bibinfo {author} {\bibfnamefont {M.}~\bibnamefont {Heurs}}, \bibinfo
  {author} {\bibfnamefont {B.}~\bibnamefont {Willke}},\ and\ \bibinfo {author}
  {\bibfnamefont {K.}~\bibnamefont {Danzmann}},\ }\bibfield  {title} {\bibinfo
  {title} {Laser power stabilization for second-generation gravitational wave
  detectors},\ }\href {https://doi.org/10.1364/OL.31.002000} {\bibfield
  {journal} {\bibinfo  {journal} {Opt. Lett.}\ }\textbf {\bibinfo {volume}
  {31}},\ \bibinfo {pages} {2000} (\bibinfo {year} {2006})}\BibitemShut
  {NoStop}%
\bibitem [{\citenamefont {Kwee}\ \emph {et~al.}(2008)\citenamefont {Kwee},
  \citenamefont {Willke},\ and\ \citenamefont {Danzmann}}]{Kwee2008}%
  \BibitemOpen
  \bibfield  {author} {\bibinfo {author} {\bibfnamefont {P.}~\bibnamefont
  {Kwee}}, \bibinfo {author} {\bibfnamefont {B.}~\bibnamefont {Willke}},\ and\
  \bibinfo {author} {\bibfnamefont {K.}~\bibnamefont {Danzmann}},\ }\bibfield
  {title} {\bibinfo {title} {Optical ac coupling to overcome limitations in the
  detection of optical power fluctuations},\ }\href
  {https://doi.org/10.1364/OL.33.001509} {\bibfield  {journal} {\bibinfo
  {journal} {Opt. Lett.}\ }\textbf {\bibinfo {volume} {33}},\ \bibinfo {pages}
  {1509} (\bibinfo {year} {2008})}\BibitemShut {NoStop}%
\bibitem [{\citenamefont {Kwee}\ \emph {et~al.}(2009)\citenamefont {Kwee},
  \citenamefont {Willke},\ and\ \citenamefont {Danzmann}}]{Kwee2009}%
  \BibitemOpen
  \bibfield  {author} {\bibinfo {author} {\bibfnamefont {P.}~\bibnamefont
  {Kwee}}, \bibinfo {author} {\bibfnamefont {B.}~\bibnamefont {Willke}},\ and\
  \bibinfo {author} {\bibfnamefont {K.}~\bibnamefont {Danzmann}},\ }\bibfield
  {title} {\bibinfo {title} {Shot-noise-limited laser power stabilization with
  a high-power photodiode array},\ }\href
  {https://doi.org/10.1364/OL.34.002912} {\bibfield  {journal} {\bibinfo
  {journal} {Opt. Lett.}\ }\textbf {\bibinfo {volume} {34}},\ \bibinfo {pages}
  {2912} (\bibinfo {year} {2009})}\BibitemShut {NoStop}%
\bibitem [{\citenamefont {Kwee}\ \emph {et~al.}(2012)\citenamefont {Kwee},
  \citenamefont {Bogan}, \citenamefont {Danzmann}, \citenamefont {Frede},
  \citenamefont {Kim}, \citenamefont {King}, \citenamefont {P\"{o}ld},
  \citenamefont {Puncken}, \citenamefont {Savage}, \citenamefont {Seifert},
  \citenamefont {Wessels}, \citenamefont {Winkelmann},\ and\ \citenamefont
  {Willke}}]{Kwee2012}%
  \BibitemOpen
  \bibfield  {author} {\bibinfo {author} {\bibfnamefont {P.}~\bibnamefont
  {Kwee}}, \bibinfo {author} {\bibfnamefont {C.}~\bibnamefont {Bogan}},
  \bibinfo {author} {\bibfnamefont {K.}~\bibnamefont {Danzmann}}, \bibinfo
  {author} {\bibfnamefont {M.}~\bibnamefont {Frede}}, \bibinfo {author}
  {\bibfnamefont {H.}~\bibnamefont {Kim}}, \bibinfo {author} {\bibfnamefont
  {P.}~\bibnamefont {King}}, \bibinfo {author} {\bibfnamefont {J.}~\bibnamefont
  {P\"{o}ld}}, \bibinfo {author} {\bibfnamefont {O.}~\bibnamefont {Puncken}},
  \bibinfo {author} {\bibfnamefont {R.~L.}\ \bibnamefont {Savage}}, \bibinfo
  {author} {\bibfnamefont {F.}~\bibnamefont {Seifert}}, \bibinfo {author}
  {\bibfnamefont {P.}~\bibnamefont {Wessels}}, \bibinfo {author} {\bibfnamefont
  {L.}~\bibnamefont {Winkelmann}},\ and\ \bibinfo {author} {\bibfnamefont
  {B.}~\bibnamefont {Willke}},\ }\bibfield  {title} {\bibinfo {title}
  {Stabilized high-power laser system for the gravitational wave detector
  advanced ligo},\ }\href {https://doi.org/10.1364/OE.20.010617} {\bibfield
  {journal} {\bibinfo  {journal} {Opt. Express}\ }\textbf {\bibinfo {volume}
  {20}},\ \bibinfo {pages} {10617} (\bibinfo {year} {2012})}\BibitemShut
  {NoStop}%
\bibitem [{\citenamefont {Junker}\ \emph {et~al.}(2017)\citenamefont {Junker},
  \citenamefont {Oppermann},\ and\ \citenamefont {Willke}}]{Junker2017}%
  \BibitemOpen
  \bibfield  {author} {\bibinfo {author} {\bibfnamefont {J.}~\bibnamefont
  {Junker}}, \bibinfo {author} {\bibfnamefont {P.}~\bibnamefont {Oppermann}},\
  and\ \bibinfo {author} {\bibfnamefont {B.}~\bibnamefont {Willke}},\
  }\bibfield  {title} {\bibinfo {title} {Shot-noise-limited laser power
  stabilization for the aei 10m prototype interferometer},\ }\href
  {https://doi.org/10.1364/OL.42.000755} {\bibfield  {journal} {\bibinfo
  {journal} {Opt. Lett.}\ }\textbf {\bibinfo {volume} {42}},\ \bibinfo {pages}
  {755} (\bibinfo {year} {2017})}\BibitemShut {NoStop}%
\bibitem [{\citenamefont {Vahlbruch}\ \emph {et~al.}(2018)\citenamefont
  {Vahlbruch}, \citenamefont {Wilken}, \citenamefont {Mehmet},\ and\
  \citenamefont {Willke}}]{Vahlbruch2018}%
  \BibitemOpen
  \bibfield  {author} {\bibinfo {author} {\bibfnamefont {H.}~\bibnamefont
  {Vahlbruch}}, \bibinfo {author} {\bibfnamefont {D.}~\bibnamefont {Wilken}},
  \bibinfo {author} {\bibfnamefont {M.}~\bibnamefont {Mehmet}},\ and\ \bibinfo
  {author} {\bibfnamefont {B.}~\bibnamefont {Willke}},\ }\bibfield  {title}
  {\bibinfo {title} {Laser power stabilization beyond the shot noise limit
  using squeezed light},\ }\href
  {https://doi.org/10.1103/PhysRevLett.121.173601} {\bibfield  {journal}
  {\bibinfo  {journal} {Phys. Rev. Lett.}\ }\textbf {\bibinfo {volume} {121}},\
  \bibinfo {pages} {173601} (\bibinfo {year} {2018})}\BibitemShut {NoStop}%
\bibitem [{\citenamefont {Wissel}\ \emph {et~al.}(2022)\citenamefont {Wissel},
  \citenamefont {Wittchen}, \citenamefont {Schwarze}, \citenamefont {Hewitson},
  \citenamefont {Heinzel},\ and\ \citenamefont {Halloin}}]{LISA2022}%
  \BibitemOpen
  \bibfield  {author} {\bibinfo {author} {\bibfnamefont {L.}~\bibnamefont
  {Wissel}}, \bibinfo {author} {\bibfnamefont {A.}~\bibnamefont {Wittchen}},
  \bibinfo {author} {\bibfnamefont {T.~S.}\ \bibnamefont {Schwarze}}, \bibinfo
  {author} {\bibfnamefont {M.}~\bibnamefont {Hewitson}}, \bibinfo {author}
  {\bibfnamefont {G.}~\bibnamefont {Heinzel}},\ and\ \bibinfo {author}
  {\bibfnamefont {H.}~\bibnamefont {Halloin}},\ }\bibfield  {title} {\bibinfo
  {title} {Relative-intensity-noise coupling in heterodyne interferometers},\
  }\href {https://doi.org/10.1103/PhysRevApplied.17.024025} {\bibfield
  {journal} {\bibinfo  {journal} {Phys. Rev. Appl.}\ }\textbf {\bibinfo
  {volume} {17}},\ \bibinfo {pages} {024025} (\bibinfo {year}
  {2022})}\BibitemShut {NoStop}%
\bibitem [{\citenamefont {Wissel}\ \emph {et~al.}(2023)\citenamefont {Wissel},
  \citenamefont {Hartwig}, \citenamefont {Bayle}, \citenamefont {Staab},
  \citenamefont {Fitzsimons}, \citenamefont {Hewitson},\ and\ \citenamefont
  {Heinzel}}]{LISA2023}%
  \BibitemOpen
  \bibfield  {author} {\bibinfo {author} {\bibfnamefont {L.}~\bibnamefont
  {Wissel}}, \bibinfo {author} {\bibfnamefont {O.}~\bibnamefont {Hartwig}},
  \bibinfo {author} {\bibfnamefont {J.}~\bibnamefont {Bayle}}, \bibinfo
  {author} {\bibfnamefont {M.}~\bibnamefont {Staab}}, \bibinfo {author}
  {\bibfnamefont {E.}~\bibnamefont {Fitzsimons}}, \bibinfo {author}
  {\bibfnamefont {M.}~\bibnamefont {Hewitson}},\ and\ \bibinfo {author}
  {\bibfnamefont {G.}~\bibnamefont {Heinzel}},\ }\bibfield  {title} {\bibinfo
  {title} {Influence of laser relative-intensity noise on the laser
  interferometer space antenna},\ }\href
  {https://doi.org/10.1103/PhysRevApplied.20.014016} {\bibfield  {journal}
  {\bibinfo  {journal} {Phys. Rev. Appl.}\ }\textbf {\bibinfo {volume} {20}},\
  \bibinfo {pages} {014016} (\bibinfo {year} {2023})}\BibitemShut {NoStop}%
\bibitem [{\citenamefont {Wysocki}\ \emph {et~al.}(1994)\citenamefont
  {Wysocki}, \citenamefont {Digonnet}, \citenamefont {Kim},\ and\ \citenamefont
  {Shaw}}]{FOG1994}%
  \BibitemOpen
  \bibfield  {author} {\bibinfo {author} {\bibfnamefont {P.}~\bibnamefont
  {Wysocki}}, \bibinfo {author} {\bibfnamefont {M.}~\bibnamefont {Digonnet}},
  \bibinfo {author} {\bibfnamefont {B.}~\bibnamefont {Kim}},\ and\ \bibinfo
  {author} {\bibfnamefont {H.}~\bibnamefont {Shaw}},\ }\bibfield  {title}
  {\bibinfo {title} {Characteristics of erbium-doped superfluorescent fiber
  sources for interferometric sensor applications},\ }\href
  {https://doi.org/10.1109/50.285318} {\bibfield  {journal} {\bibinfo
  {journal} {Journal of Lightwave Technology}\ }\textbf {\bibinfo {volume}
  {12}},\ \bibinfo {pages} {550} (\bibinfo {year} {1994})}\BibitemShut
  {NoStop}%
\bibitem [{\citenamefont {Rovera}\ \emph {et~al.}(2023)\citenamefont {Rovera},
  \citenamefont {Tancau}, \citenamefont {Boetti}, \citenamefont {Dalla~Vedova},
  \citenamefont {Maggiore},\ and\ \citenamefont {Janner}}]{FOS2023}%
  \BibitemOpen
  \bibfield  {author} {\bibinfo {author} {\bibfnamefont {A.}~\bibnamefont
  {Rovera}}, \bibinfo {author} {\bibfnamefont {A.}~\bibnamefont {Tancau}},
  \bibinfo {author} {\bibfnamefont {N.}~\bibnamefont {Boetti}}, \bibinfo
  {author} {\bibfnamefont {M.~D.~L.}\ \bibnamefont {Dalla~Vedova}}, \bibinfo
  {author} {\bibfnamefont {P.}~\bibnamefont {Maggiore}},\ and\ \bibinfo
  {author} {\bibfnamefont {D.}~\bibnamefont {Janner}},\ }\bibfield  {title}
  {\bibinfo {title} {Fiber optic sensors for harsh and high radiation
  environments in aerospace applications},\ }\href@noop {} {\bibfield
  {journal} {\bibinfo  {journal} {Sensors}\ }\textbf {\bibinfo {volume} {23}}
  (\bibinfo {year} {2023})}\BibitemShut {NoStop}%
\bibitem [{\citenamefont {{Robertson}}\ \emph {et~al.}(1986)\citenamefont
  {{Robertson}}, \citenamefont {{Hoggan}}, \citenamefont {{Mangan}},\ and\
  \citenamefont {{Hough}}}]{Robertson1986}%
  \BibitemOpen
  \bibfield  {author} {\bibinfo {author} {\bibfnamefont {N.~A.}\ \bibnamefont
  {{Robertson}}}, \bibinfo {author} {\bibfnamefont {S.}~\bibnamefont
  {{Hoggan}}}, \bibinfo {author} {\bibfnamefont {J.~B.}\ \bibnamefont
  {{Mangan}}},\ and\ \bibinfo {author} {\bibfnamefont {J.}~\bibnamefont
  {{Hough}}},\ }\bibfield  {title} {\bibinfo {title} {{Intensity stabilisation
  of an argon laser using an electro-optic modulator - Performance and
  limitations}},\ }\href@noop {} {\bibfield  {journal} {\bibinfo  {journal}
  {Applied Physics B Photophysics Laser Chemistry}\ }\textbf {\bibinfo {volume}
  {39}},\ \bibinfo {pages} {149} (\bibinfo {year} {1986})}\BibitemShut
  {NoStop}%
\bibitem [{\citenamefont {Tran}\ and\ \citenamefont {Furlan}(1993)}]{Tran1993}%
  \BibitemOpen
  \bibfield  {author} {\bibinfo {author} {\bibfnamefont {C.~D.}\ \bibnamefont
  {Tran}}\ and\ \bibinfo {author} {\bibfnamefont {R.~J.}\ \bibnamefont
  {Furlan}},\ }\bibfield  {title} {\bibinfo {title} {Indirect amplitude
  stabilization of a tunable laser through control of the intensity of a pump
  laser by an electro-optic modulator},\ }\href
  {https://opg.optica.org/as/abstract.cfm?URI=as-47-2-235} {\bibfield
  {journal} {\bibinfo  {journal} {Appl. Spectrosc.}\ }\textbf {\bibinfo
  {volume} {47}},\ \bibinfo {pages} {235} (\bibinfo {year} {1993})}\BibitemShut
  {NoStop}%
\bibitem [{\citenamefont {Kaufer}\ and\ \citenamefont
  {Willke}(2019)}]{Kaufer2019}%
  \BibitemOpen
  \bibfield  {author} {\bibinfo {author} {\bibfnamefont {S.}~\bibnamefont
  {Kaufer}}\ and\ \bibinfo {author} {\bibfnamefont {B.}~\bibnamefont
  {Willke}},\ }\bibfield  {title} {\bibinfo {title} {Optical ac coupling power
  stabilization at frequencies close to the gravitational wave detection
  band},\ }\href {https://doi.org/10.1364/OL.44.001916} {\bibfield  {journal}
  {\bibinfo  {journal} {Opt. Lett.}\ }\textbf {\bibinfo {volume} {44}},\
  \bibinfo {pages} {1916} (\bibinfo {year} {2019})}\BibitemShut {NoStop}%
\bibitem [{\citenamefont {Nery}\ \emph {et~al.}(2021)\citenamefont {Nery},
  \citenamefont {Venneberg}, \citenamefont {Aggarwal}, \citenamefont {Cole},
  \citenamefont {Corbitt}, \citenamefont {Cripe}, \citenamefont {Lanza},\ and\
  \citenamefont {Willke}}]{TradNery2021}%
  \BibitemOpen
  \bibfield  {author} {\bibinfo {author} {\bibfnamefont {M.~T.}\ \bibnamefont
  {Nery}}, \bibinfo {author} {\bibfnamefont {J.~R.}\ \bibnamefont {Venneberg}},
  \bibinfo {author} {\bibfnamefont {N.}~\bibnamefont {Aggarwal}}, \bibinfo
  {author} {\bibfnamefont {G.~D.}\ \bibnamefont {Cole}}, \bibinfo {author}
  {\bibfnamefont {T.}~\bibnamefont {Corbitt}}, \bibinfo {author} {\bibfnamefont
  {J.}~\bibnamefont {Cripe}}, \bibinfo {author} {\bibfnamefont
  {R.}~\bibnamefont {Lanza}},\ and\ \bibinfo {author} {\bibfnamefont
  {B.}~\bibnamefont {Willke}},\ }\bibfield  {title} {\bibinfo {title} {Laser
  power stabilization via radiation pressure},\ }\href
  {https://doi.org/10.1364/OL.422614} {\bibfield  {journal} {\bibinfo
  {journal} {Opt. Lett.}\ }\textbf {\bibinfo {volume} {46}},\ \bibinfo {pages}
  {1946} (\bibinfo {year} {2021})}\BibitemShut {NoStop}%
\bibitem [{\citenamefont {Guan}\ \emph {et~al.}(2021)\citenamefont {Guan},
  \citenamefont {Zhang}, \citenamefont {Shang}, \citenamefont {Pan},
  \citenamefont {He}, \citenamefont {Pan},\ and\ \citenamefont
  {Chen}}]{Guan2021}%
  \BibitemOpen
  \bibfield  {author} {\bibinfo {author} {\bibfnamefont {X.}~\bibnamefont
  {Guan}}, \bibinfo {author} {\bibfnamefont {T.}~\bibnamefont {Zhang}},
  \bibinfo {author} {\bibfnamefont {H.}~\bibnamefont {Shang}}, \bibinfo
  {author} {\bibfnamefont {D.}~\bibnamefont {Pan}}, \bibinfo {author}
  {\bibfnamefont {J.}~\bibnamefont {He}}, \bibinfo {author} {\bibfnamefont
  {J.}~\bibnamefont {Pan}},\ and\ \bibinfo {author} {\bibfnamefont
  {J.}~\bibnamefont {Chen}},\ }\bibfield  {title} {\bibinfo {title} {Improving
  laser power stability with a photosensitive lens},\ }\href
  {https://doi.org/10.1063/5.0048119} {\bibfield  {journal} {\bibinfo
  {journal} {Review of Scientific Instruments}\ }\textbf {\bibinfo {volume}
  {92}},\ \bibinfo {pages} {083003} (\bibinfo {year} {2021})}\BibitemShut
  {NoStop}%
\bibitem [{\citenamefont {Jie}\ \emph {et~al.}(2021)\citenamefont {Jie},
  \citenamefont {Guangyao}, \citenamefont {Guochao}, \citenamefont {Yaning},
  \citenamefont {Mei}, \citenamefont {Qixue}, \citenamefont {Lingxiao},
  \citenamefont {Xinghui}, \citenamefont {Shuhua},\ and\ \citenamefont
  {Jun}}]{Wang2021}%
  \BibitemOpen
  \bibfield  {author} {\bibinfo {author} {\bibfnamefont {W.}~\bibnamefont
  {Jie}}, \bibinfo {author} {\bibfnamefont {H.}~\bibnamefont {Guangyao}},
  \bibinfo {author} {\bibfnamefont {W.}~\bibnamefont {Guochao}}, \bibinfo
  {author} {\bibfnamefont {W.}~\bibnamefont {Yaning}}, \bibinfo {author}
  {\bibfnamefont {H.}~\bibnamefont {Mei}}, \bibinfo {author} {\bibfnamefont
  {L.}~\bibnamefont {Qixue}}, \bibinfo {author} {\bibfnamefont
  {Z.}~\bibnamefont {Lingxiao}}, \bibinfo {author} {\bibfnamefont
  {L.}~\bibnamefont {Xinghui}}, \bibinfo {author} {\bibfnamefont
  {Y.}~\bibnamefont {Shuhua}},\ and\ \bibinfo {author} {\bibfnamefont
  {Y.}~\bibnamefont {Jun}},\ }\bibfield  {title} {\bibinfo {title}
  {One-thousandth-level laser power stabilization based on optical feedback
  from a well-designed high-split-ratio and nonpolarized beam splitter},\
  }\href {https://doi.org/10.1364/AO.431994} {\bibfield  {journal} {\bibinfo
  {journal} {Appl. Opt.}\ }\textbf {\bibinfo {volume} {60}},\ \bibinfo {pages}
  {7798} (\bibinfo {year} {2021})}\BibitemShut {NoStop}%
\bibitem [{\citenamefont {Tricot}\ \emph {et~al.}(2018)\citenamefont {Tricot},
  \citenamefont {Phung}, \citenamefont {Lours}, \citenamefont {Guérandel},\
  and\ \citenamefont {de~Clercq}}]{Tricot2018}%
  \BibitemOpen
  \bibfield  {author} {\bibinfo {author} {\bibfnamefont {F.}~\bibnamefont
  {Tricot}}, \bibinfo {author} {\bibfnamefont {D.~H.}\ \bibnamefont {Phung}},
  \bibinfo {author} {\bibfnamefont {M.}~\bibnamefont {Lours}}, \bibinfo
  {author} {\bibfnamefont {S.}~\bibnamefont {Guérandel}},\ and\ \bibinfo
  {author} {\bibfnamefont {E.}~\bibnamefont {de~Clercq}},\ }\bibfield  {title}
  {\bibinfo {title} {Power stabilization of a diode laser with an acousto-optic
  modulator},\ }\href {https://doi.org/10.1063/1.5046852} {\bibfield  {journal}
  {\bibinfo  {journal} {Review of Scientific Instruments}\ }\textbf {\bibinfo
  {volume} {89}},\ \bibinfo {pages} {113112} (\bibinfo {year}
  {2018})}\BibitemShut {NoStop}%
\bibitem [{\citenamefont {Kobayashi}\ \emph {et~al.}(2006)\citenamefont
  {Kobayashi}, \citenamefont {Izumi}, \citenamefont {Kumakura},\ and\
  \citenamefont {Takahashi}}]{Kobayashi2006}%
  \BibitemOpen
  \bibfield  {author} {\bibinfo {author} {\bibfnamefont {J.}~\bibnamefont
  {Kobayashi}}, \bibinfo {author} {\bibfnamefont {Y.}~\bibnamefont {Izumi}},
  \bibinfo {author} {\bibfnamefont {M.}~\bibnamefont {Kumakura}},\ and\
  \bibinfo {author} {\bibfnamefont {Y.}~\bibnamefont {Takahashi}},\ }\bibfield
  {title} {\bibinfo {title} {Stable all-optical formation of bose–einstein
  condensate using pointing-stabilized optical trapping beams},\ }\href
  {https://doi.org/10.1007/S00340-006-2140-2} {\bibfield  {journal} {\bibinfo
  {journal} {Applied Physics B}\ }\textbf {\bibinfo {volume} {83}},\ \bibinfo
  {pages} {21} (\bibinfo {year} {2006})}\BibitemShut {NoStop}%
\bibitem [{\citenamefont {Zhang}\ \emph {et~al.}(2019)\citenamefont {Zhang},
  \citenamefont {Chen}, \citenamefont {Fang}, \citenamefont {Wang},
  \citenamefont {Li},\ and\ \citenamefont {Luo}}]{Zhang2019}%
  \BibitemOpen
  \bibfield  {author} {\bibinfo {author} {\bibfnamefont {X.}~\bibnamefont
  {Zhang}}, \bibinfo {author} {\bibfnamefont {Y.}~\bibnamefont {Chen}},
  \bibinfo {author} {\bibfnamefont {J.}~\bibnamefont {Fang}}, \bibinfo {author}
  {\bibfnamefont {T.}~\bibnamefont {Wang}}, \bibinfo {author} {\bibfnamefont
  {J.}~\bibnamefont {Li}},\ and\ \bibinfo {author} {\bibfnamefont
  {L.}~\bibnamefont {Luo}},\ }\bibfield  {title} {\bibinfo {title} {Beam
  pointing stabilization of an acousto-optic modulator with thermal control},\
  }\href {https://doi.org/10.1364/OE.27.011503} {\bibfield  {journal} {\bibinfo
   {journal} {Opt. Express}\ }\textbf {\bibinfo {volume} {27}},\ \bibinfo
  {pages} {11503} (\bibinfo {year} {2019})}\BibitemShut {NoStop}%
\bibitem [{\citenamefont {Deng}\ and\ \citenamefont {Shen}(2023)}]{Deng2023}%
  \BibitemOpen
  \bibfield  {author} {\bibinfo {author} {\bibfnamefont {S.}~\bibnamefont
  {Deng}}\ and\ \bibinfo {author} {\bibfnamefont {H.}~\bibnamefont {Shen}},\
  }\bibfield  {title} {\bibinfo {title} {Influence of acousto-optic frequency
  shifter's thermal-induced birefringence on laser frequency-shifted feedback
  system},\ }\href
  {https://doi.org/https://doi.org/10.1016/j.optlaseng.2022.107290} {\bibfield
  {journal} {\bibinfo  {journal} {Optics and Lasers in Engineering}\ }\textbf
  {\bibinfo {volume} {160}},\ \bibinfo {pages} {107290} (\bibinfo {year}
  {2023})}\BibitemShut {NoStop}%
\end{thebibliography}
%
\end{document}